# Redescription and validation of *Bothriechis supraciliaris* (Serpentes: Viperidae)

Alejandro Solórzano [1], Luis D. Gómez [2], Julián Monge-Nájera [3], Brian I. Crother [4]

[1] Sepentario Nacional. Ave. 1. c. 9/11. San José, Costa Rica, fax: (506) 233-5520.
[2] Jardín Wilson, Organización para Estudios Tropicales, Las Cruces, Coto Brus, Costa Rica.
[3] Centro de lnvestigación Académica, Universidad Estatal a Distancia, Sabanilla M. Oca, San José, Costa Rica. Mailing address: Biologia Tropical, Universidad de Costa Rica, 2060 San José. Costa Rica. Telefax (506)2075550; julianm@cariari.ucr.ac.cr
[4] Dept. Biological Sciences. SE Louisiana University, Hammond, LA. USA.

**Abstract:** The populations of pitvipers from south western Costa Rica, have traditionally been identified as *Bothriechis schlegelii* (Berthold). However, in 1954 E. H. Taylor described one specimen from the area as a new subspecies, *B. schlegelii supraciliaris*. Werman returned *supraciliaris* to synonymy with *schlegelii* four decades later. However, morphometry and color pattern in a SW Costa Rica population (25 specimens) differ from those of specimens (N=57) from other parts of Costa Rica and from descriptions of South American specimens. Here the epithet *Bothriechis schlegelii supraciliaris* Taylor 1954. is reestablished as a valid taxon and elevated to specific rank as *B. supraciliaris* stat.nov. It is closely related to *B. schlegelii* from which it differs by its color patterns based on a uniform ground color with polymorphic dorsal designs, and its lower counts of ventral and caudal scales.

In 1954 Edward H. Taylor described a specimen from Valle del General (San José province, South West Costa Rica), as a subspecies of *Bothrops schlegelii* (Berthold), which he named *Bothrops schlegelii supraciliaris.* Thirty years later, Werman (1984) re-examined the type and concluded that it did not warrant subspecific status. Stuart (1963) had agreed with Taylor that the specimen was not typical *B. schlegelii,* and was followed by Peters & Orejas-Miranda (1970) but not by Bolaños (1984), Savage (1973, 1980), Taylor *et al.* (1974), or Villa (1984).

Wilson & Meyer (1982, 1985) examined Taylor's type and again based on that single specimen, considered it as a slightly aberrant form of *B. schlegelii.* However, a series of specimens collected in the mid-elevations of Coto Brus, increased the sample size to 26 and showed remarkable deviations from the typical *schlegelii* patterns of coloration and prompted a review. Based on coloration and scale counts we propose that Taylor's taxon is a discrete entity and should be accorded specific rank.

## MATERIALS AND METHODS

Fifty-seven specimens of typical *Bothriechis schlegelii* (sensu lato), from various parts of Costa Rica and twenty-five specimens collected in the forest reserve of the Las Cruces Biological Station, Coto Brus, Puntarenas, at 1000-1400 m a.s.l., were compared (numbers and localities in Appendix 1) for scale counts and for color patterns of live specimens.

## RESULTS

**Scale counts:** Coto Brus specimens (Table 1 and Fig. 4): dorsal rows 21-23; supralabials 7-11; infralabials 10-12; interoculars 6-9; preoculars 2-4; postoculars 2- 4; 1 elongated subocular; 1-2 (rarely 3) scales between subocular and supralabials; anal scale entire. Dorsal scales keeled, except for paraventrals. Dorsocaudals keeled.

**Color pattern:** Specimens from Coto Brus have consecutive, well-defined blotches of variable shapes (circular, rhomboidal, disjunct circles and bands, transverse banding) along center of dorsum, ranging from coffee brown or dark green to rusty brown or reddish maroon. These patterns visibly differ from the rather diffuse coloration of all specimens known from elsewhere in Costa Rica (Figs. 1- 3). Additionally, the ground color of the Coto Brus specimens, albeit highly variable in hues and shades, is almost always uniform and lacks secondary pigments, while the dorsal ground coloration of specimens from elsewhere in the country is highlighted by a variety of dots, spots or bars of secondary pigmentation (Figs. 1 and 3). Furthermore, the "oropel" morph of typical *B. schlegelii,* yellowish to golden orange, so well known from other Costa Rican sites is extremely rare in the General-Coto Brus area, where only one specimen (immature male) was seen in ten years of observations by one of the authors (L.D.G.).

Ventral coloration outside Coto Brus is equally enriched by scattered dots and spots, while in specimens from Coto Brus the ventral aspect is devoid of pigments at least in the anterior two thirds of the body length (Fig. 3), the caudal portion rarely has diffuse, dark, purplish brown hues mostly visible in preserved material.

Lepidosis: Specimens from Coto Brus and those from elsewhere in Costa Rica differ in number of ventral and caudal scales (Fig. 4). There are no significant differences in length and number of dorsal scales (Fig. 4).

TABLE 1

*Ventral and caudal scale counts in* Bothriechis schlegelii *and* B. supraciliaris. *Mean in bold to facilitate comparisons*

|  |  | *B. schlegelii* | | | | *B. supraciliaris* | | | |
|---|---|---|---|---|---|---|---|---|---|
|  |  | **Mean** | S.D. | Range | N | **Mean** | S.D. | Range | N |
| Ventrals | Males | **160.1** | 2.44 | 153-167 | 9 | **147.3** | 2.22 | 145-150 | 4 |
|  | Females | **156.8** | 4.75 | 146-167 | 48 | **143.9** | 2.13 | 141-148 | 14 |
| Caudals | Males | **56.8** | 2.44 | 53-61 | 9 | **47.3 50.2** | 2.63 | 48-54 | 4 |
|  | Females | **51.6** | 4.34 | 33-61 | 43 |  | 2.06 | 45-52 | 14 |

*S.D*. Standard Deviation. Range (Minimum-Maximum). *N* Sample size.

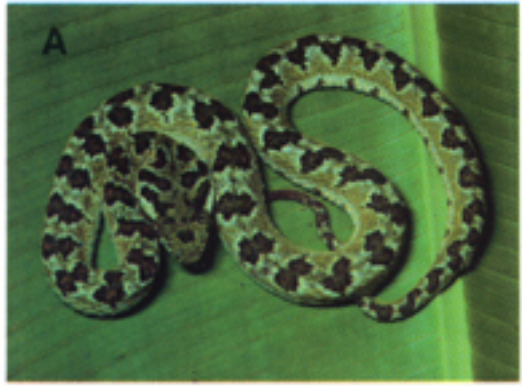
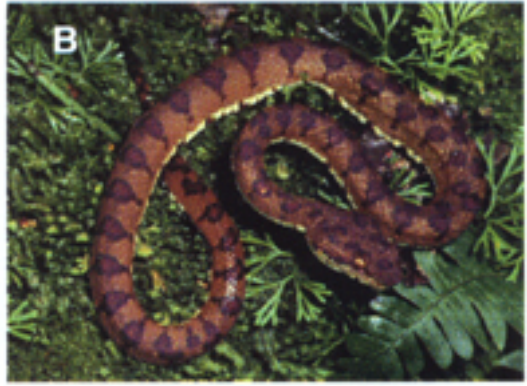
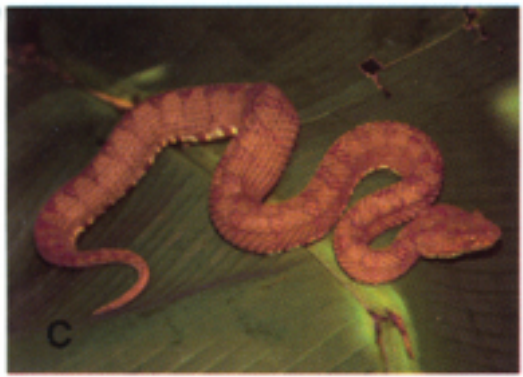
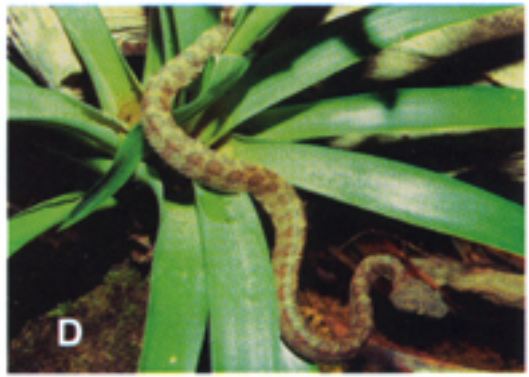
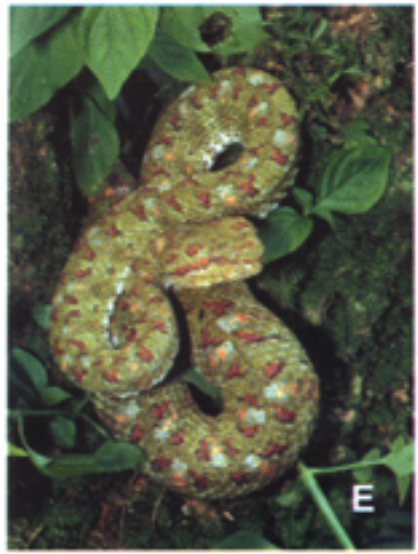

Fig. 1. Variation of color patterns. A-D B. supraciliaris, E B. schlegelii, all Costa Rica (photographs: A, B: M. & P. Fogden: C, D, E: A. Solórzano)

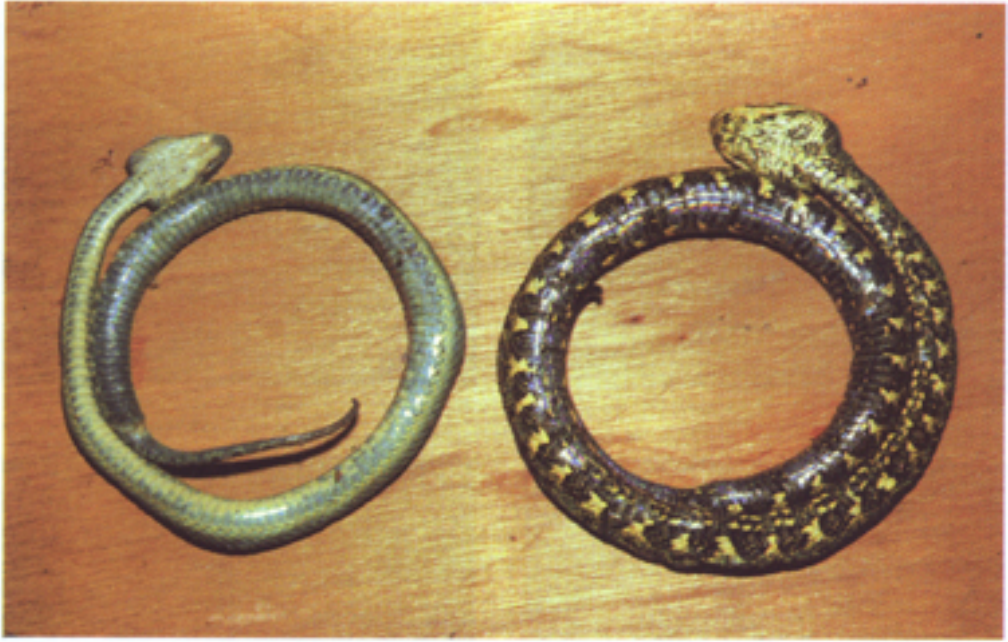

Fig. 2. Ventral view of *B. supraciliaris* (left) and of *B. schlegelii*

Based on these results, we conclude that *Bothriechis supraciliaris* warrants specific recognition and it is here redescribed:

*Bothriechis supraciliaris* (Taylor 1954), Fig. 1.

**Diagnosis:** Closely related to *B. schlegelii* from which it differs by its color patterns based on a uniform ground color with polymorphic, dorsal designs and its lower counts of ventral and caudal scales.

**Description:** Pitviper of small to moderate site (maximum 600 mm), prehensile tail and strongly triangular outline of head and, from Taylor's original description: "Front of snout rounding, covered by about 70 strongly keeled scales anterior to level of the supraoculars; rostral approximately as wide as high, not visible above: on canthus a tiny median scale above rostral; each undivided nasal bordered above by three scales with free outer edges; loreal separated from the upper preocular by one small scale; these three scales bordered above by five scales, three of which have the keels strongly elevated into soft, flattened "spines" and forming a line with the two elevated supraorbital spines: the supraoculars divided, each bearing one or two small, spine-like outer projections: supraoculars separated from the eyes by a row of ten small granular scales: the next row consists of about five scales, two of which are the elevated supraorbital spines, this row separated from orbit by eight or nine small granules, two or three small postoculars: a long narrow subocular runs under orbit to or very close to the lower preocular: three preoculars, the median and lower border pit: second labial separated from lower preocular by small scale, broadly entering pit and with a scale bordering it above forms the anterior border of pit: the second labial has no partial suture entering from its anterior edge; three to five small lower loreals (prefoveals); 9-10 supralabials, the first continuously bordering the nasal: the third, fourth, and sixth separated from the subocular by three scale rows; infralabials 12-12, only two of which touch the first chinshields; first pair of chinshields much larger than the three following pairs, but distinctly smaller but perhaps a little wider than the first pair of labials; temporal scales keeled strongly, the keels compressed and elevated; about 56 scales across head at angle of jaw; no scales between nasal and rostral. Scale formula: *25,* 23, 23, 23, 19, 19; ventrals (counting from first widened scale), 146, caudals 46, anal single; the terminal scute of tail rather large; the tail a tip with six rows of scales; 15 in the postanal region; all scales except outer row, and scales on underside of head, keeled."

Our scale counts of the holotype agree with Taylor's, except that the caudals are 47. There is a marked sexual dimorphism: the females are longer and thicker (pers. observ. of 14 females and 4 males) as in *B. schlegelii.*

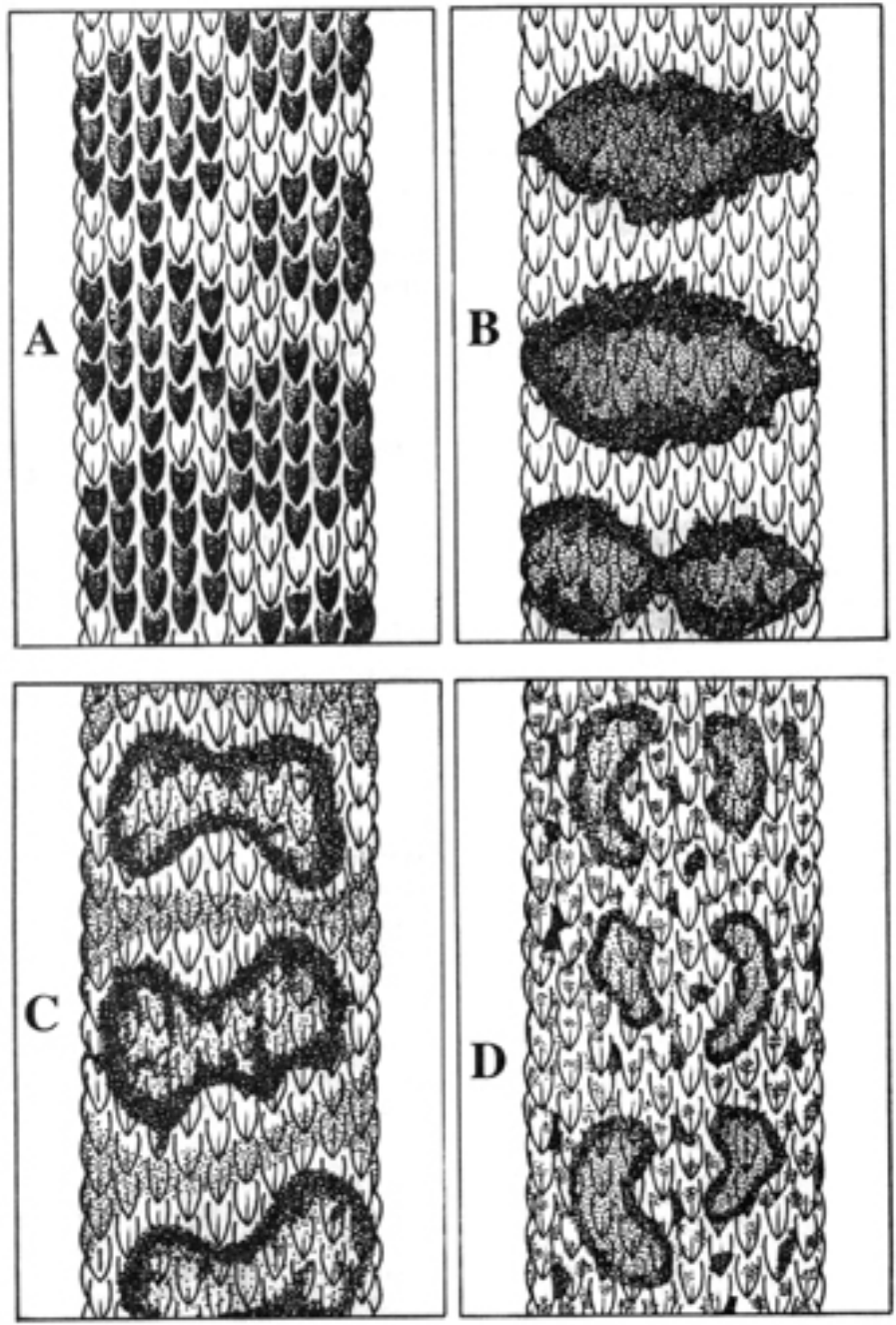

Fig. 3. Dorsal scale pattern in B. supraciliaris (A-C) and of B. Schlegelii

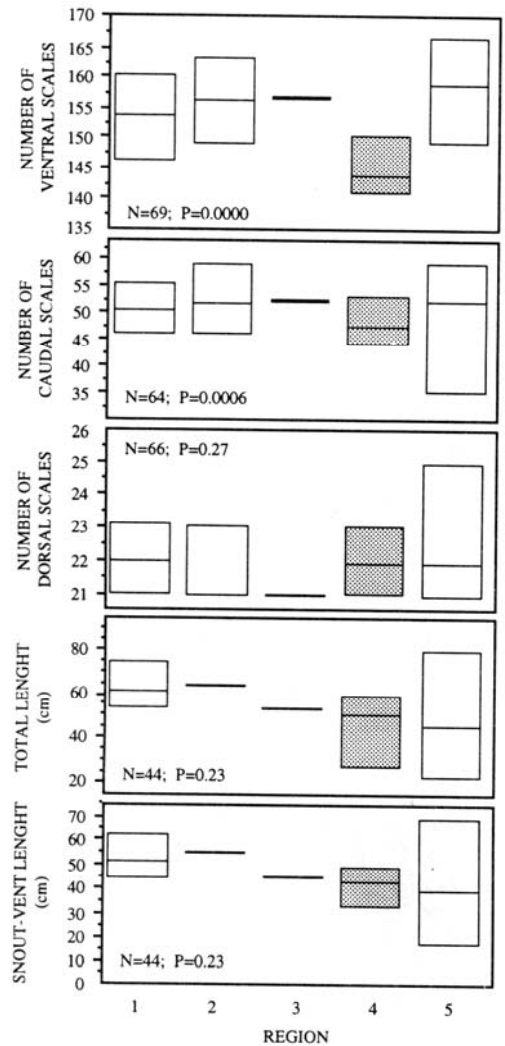

Fig. 4. Descriptive morphometry of populations previously considered *"B. schlegelii"* in five regions of Costa Rica. The country was divided in five sections: 1 North Pacific, 2 Central Pacific, 3 South Pacific, 4 San Vito de Corn Brus and San Isidro de Perez ZelecJOn and *5* Caribbean. Section 4 differs significantly from the other areas, which are statistically similar among themselves (Kruskal-Wallis ANOVA and Tukey test, p<0.05). Values are: maximum, mean and minimum. Sample sizes vary because some specimens were damaged.

**Natural history:** *B. supraciliaris* may tend to spend more time on the ground than other palm or eyelash pitvipers: 22 out of 25 were found on the forest floor. The stomach contents of four dissected specimens consisted mostly of small rodents of the forest floor.

Specimens collected in resting positions were found curled in a loose circle and mostly on horizontal, wide surfaces, while typical *schlegelii* often droop and loop on forks of stems and branches or form a tight S-shaped pattern on almost vertical, rough-barked tree trunks.

The following specimens are proposed as neoparatypes, deposited at Museo de Zoologfa, Universidad de Costa Rica (UCR) and Collection of Vertebrates, University of Texas at Arlington (UTA): Alturas del RIo Cotdn, Coto Brus, Puntarenas Prov., 1500 m UCR-12082; Estación Biológica Las Cruces, Coto Brus, 1000-1400 m, Puntarenas Prov. UCR- 10641, 12075-12081, 13500-13502, 2276, 2277. UTA R-35 192, 35193, 35194, 35246.

DISCUSSION

Among neotropical pitvipers *Bothriechis schlegelii* is known particularly for its impressive color variation throughout its geographical range (Werman 1984, Campbell & Lamar 1989). Its usual habitat is lowland rainforests although it is also known from moderate to high elevations (Campbell & Lamar 1989, Crother *et a!.* 1992). In Costa Rica *B. schlegelii* is amply distributed in the Caribbean and southern Pacific lowlands, but the Pacific range is geographicaly isolated from the rest of the country (Crother *et al.* 1992).

Recently, the systematics of the genera *Bothrops* and *Bothriechis* has been the subject of diverse morphological and biochemical studies (Campbell & Lamar 1989, 1992, Shätti *et al.* 1990, & Kramer 1991, Cadle 1992, Crother *et al.* 1992, Werman 1992, Kuch and Freire 1995, Wüster and McCarthy 1996). Taylor's (1954) keen observation of a single specimen (Type, KU. 31997) was notable and gross differences shown by the material at hand were correct, although the main parameters used for his description were not, as he based the distinction on the aspect and morphology of some scales that often are ontogenetically variable within the species, such as the divided supraoculars and the color of the tail tip.

This study shows that, despite normal variation, specimens from Coto Brus display a notable and constant coloration divergence from the rest of the country.

Coloration of the Costa Rican taxon is equally and strongly related to some morphotypes from Colombia and Ecuador as described and illustrated in Campbell & Lamar (1989). The close resemblance and affinity of the SW Costa Rican Pacific versant *supraciliaris* to the NW South American *schlegelii* seems to support the hypothesis of an ancestral continuum of a *schlegelii* stock, a hypothesis further enhanced by the existence of other cases (Solórzano 1995), such as *Porthidium volcanicum* Solórzano and its vicar *P kinsbergii,* or by *Lachesis muta melanocephala* Solórzano & Cerdas (1986) whose color pattern and scaling is closer to the South American *L muta muta* than to *L muta stenophrys* from the Atlantic versant of Costa Rica. In contrast, Zamudio and Greene (1997) defend a greater affinity between this and *stenophiys.*

Unlike Werman (1984), we used a larger sample that has evidenced significant differences in the number of ventral and caudal scales. Allozyme comparisons of five individuals from Coto Brus, one from Upala in NE Costa Rica, and one from Ciudad Neily, *35* km SW of Coto Brus, support the differentiation (Crother, unpublished).

The geographical range of *supraciliaris* seems to be confined to mid and higher elevations (800-1700 m) (Appendix 1). For *schlegelii,* the range is ampler but mostly confined to lowland rainforests outside and within SW Costa Rica. To date, we have found no overlap of the ranges of the two taxa. There is a closer similarity in squamation between Costa Rican *supraciliaris* and Colombian *schlegelii* (see Renjifo 1979 for summary statistics of Colombian populations).

The species redescribed here is thus far only known from the Valle del General and the Coto Brus altiplano in SW Costa Rica, and its existence suggests that an exhaustive revision of the systematics of *Bothriechis schlegelii* throughout its range, from Northern Penú to Mexico, would likely reveal similarly isolated and discrete genotypes.

AKNOWLEDGMENTS
We are grateful to William W. Lamar, Mahmoud Sasa, José M. Gutiérrez and two anonymous reviewers for suggestions to greatly improve this manuscript, and to Alberto Leon S. Peters iA. & B. Orejas-Miranda. 1970. Catalogue of the for his drawings. Michael and Patricia Fogden supplied some photographs. The organization for Tropical Studies, provided logistical support for the two senior authors.

RESUMEN

Las poblaciones de toboba de pestañas o bocaracá del suroeste de Costa Rica han sido denominadas tradicionalmente *Bothriechis schlegelii supraciliaris* Taylor 1954. Sin embargo, la morfometría y el patron de coloración sugieren que es una especie aparte, que aquí se propone como *Bothriechis supraciliaris* stat.nov.

REFERENCES

Berthold, A.A. 1846. Ueber verschiedene neue oder seltene Reptilien aus New Granada und Crustacien aus China. Abh. Ges. Wiss. Gottingen 3: 3-32.

Bolaños, R. 1984. Serpientes, venenos y ofidismo en Centro América. Universidad de Costa Rica, San José, Costa Rica, 136 p.

Cadle, 1.E. 1992. Phylogenetic relationships among Vipers: Inmunological evidence. pp. 41-48. *In* J.A.Campbell & E.O. Brodie (eds.). Biology of the Pitvipers. Selva, Tyler, Texas.

Campbell, J.A. & W.W. Lamar. 1989. The venomous reptiles of Latin America. Cornell University, Ithaca, New York, 425 p.

Campbell, J.A. & W.W. Lamar. 1992. Taxonomic status of miscellaneous neotropical viperids, with the description of a new genus. Occas.Papers Mus.,Texas Tech Univ. 153: 1-31.

Crother, B.I., 1.A. Campbell & D.M. Hillis. 1992. Phylogeny and Historical Biogeography of the Palm - Pitvipers, genus *Bothriechis:* Biochemical and Morphological Evidence, p. 1-20. *In* J.A. Campbell & E.O. Brodie, Jr.(eds.) Biology of the Pitvipers. Selva, Tyler, Texas.

Kuch, U. & A. Freire. 1995. Bemerkungen zur geographischen Verbreitung und Variabilitat von Schlegels Palmen-Lanzenotter, *Bothriechis schlegelii* (BERTHOLD, 1846), in Ecuador. Herpetozoa 8: 49-58.

Peters J.A. & B. Orejas-Miranda. 1970. Catalogue of the Neotropical Squamata. Part 1. Snakes. Bull. U.S. Natl. Mus.297: 1347.

Renjifo, J.M. 1979. Systematics and distribution of Crotalid snakes in Colombia. MS Thesis, University of Kansas, Lawrence, 64 p.

Savage, J.M. 1973. A prelimary handlist of the herpetofauna of Costa Rica. Univ. S. California, Los Angeles, California, 17p.

Savage, J.M. 1980. A handlist with preliminary keys to the herpetofauna of Costa Rica. J. Allan Hancock Foundation, Los Angeles, California, 111 p.

Schätti, B. & E. Kramer. 1991. A new pitviper from Ecuador, *Bothriechis mahnerti* n. sp. Rev. Suisse Zool. 98: 9-14.

Schätti, B., E. Kramer & J.M. Touzet. 1990. Systematic remarks on a rare crotalid snake from Ecuador, *Bothriechis albocarinata* (Shreve), with some comments on the generic arrangement of arboreal neotropical pitvipers. Rev. Suisse Zool. 97: 877-885.

Solórzano, A. 1995. Una nueva especie de serpiente venenosa terrestre del género *Porthidium* (Serpentes: Viperidae), del Suroeste de Costa Rica. Rev. Biol. Trop. 42: 695-701.

Solórzano, A. & L. Cerdas. 1986. A new subspecies of the Bushmaster, *Lachesis muta,* from SW Costa Rica. J. Herpetol. 20: 463-466.

Stuart, L.C. 1963. A checklist of the herpetofauna of Guatemala. Misc. Publ. Mus. Zool. Univ. Michigan 122: 1-150.

Taylor, E.H. 1954. Further studies on tbe serpents of Costa Rica. Univ. Kansas Sci. Bull. 36: 637-801.

Taylor, R.T., A. Flores & R. Bolaños. 1974. Geographical distribution of Viperidae, Elapidae and Hydrophidae in Costa Rica. Rev. Biol. Trop. 21: 383-397.

Villa, J. D. 1984. The venomous snakes of Nicaragua: a synopsis. Contr. Milwaukee Public Museum 59: 29-31.

Werman, S.O. 1984. The taxonomic status of *Bothrops supraciliaris* Taylor. J. Herpetol. 18: 484-486.

Werman, S.O. 1992. Phylogenetic relationships of Central and South American pitvipers of tbe genus *Bothrops (sensu Latu):* Cladistic analyses of biochemical and anatomical characters. p. 21-40. *In* J.A. Campbell & E.D. Brodie Jr (rds.). Biology of the pitvipers, Selva, Tyler, Texas.

Wilson, L.D. & J.R. Meyer. 1982. The snakes of Honduras. Milwaukee Publ. Mus. Publ., Biol. Geol. 6: 1-159.

Wilson, L.D. & J.R. Meyer. 1985. The snakes of Honduras. Milwaukee Publ. Mus. Publ., Biol. Geol. 6: 1-150.

Wüster, W. & CJ. McCarty. 1996. Venomous snake systematics: lmplications for snake bite treatment and toxinology, p. 13·23. *In* C. Bon & M. Boyffon (eds.). Envenomings and their treatrnents. Foundation Mareel Mérieux, Lyon, France.

Zamudio, K.R. & H.W. Greene. 1997. Phylogeography of the bushmaster *(Lachesis muta:* Viperidae): lmplieations for neotropical biogeography, systematics, and conservation. Biol. J. Linn. Soco 62: 421-442.

APPENDIX 1

*Studied specimens of Bothriechis schlegelii (from outside SW Costa Rica; for B. supraciliaris data see final paragraph of Results)*

| UCR-Number | Collection locality | Specimen Number | Collection locality |
|---|---|---|---|
| 3309 | Osa, Puntarenas | | Reference collection, Instituto Clodomiro Picado, San José, Costa Rica. |
| 2879 | Penshurt, Limón | | |
| 3628 | Guápiles, Limón | | |
| 7161 | Comadre, Limón | | |
| 1439 | Tilarán, Guanacaste | 1- | Llano Hermoso de Puriscal, San José |
| 6098 | La Tirimbina de Sarapiquí, Heredia | 2- | Mercedes Sur de Puriscal, San José |
| 3389 | Turrubares, San José | 3- | Salitrales de Puriscal, San José |
| 10439 | Brasilia de Upala, Alajuela | 4- | Salitrales de Puriscal, San José |
| 10415 | Dos Ríos de Upala, Alajuela | 5- | Salitrales de Puriscal, San José |
| 10416 | Dos Ríos de Upala, Alajuela | 6- | Salitrales de Puriscal, San José |
| 3316 | Jesús María de Turrialba, Cartago | 7- | Salitrales de Puriscal, San José |
| 0013 | San José | 8- | Salitrales de Puriscal, San José |
| 3437 | Tilarán, Guanacaste | 9- | Tilarán, Guanacaste |
| 1430 | Puerto Viejo de Sarapiquí, Heredia | 10- | Concepción, Puriscal, San José |
| 7197 | Río Cuarto de Grecia, Alajuela | 11- | Concepción, Puriscal, San José |
| 0326 | Puerto Viejo de Sarapiquí, Heredia | 12- | Concepción, Puriscal, San José |
| 2595 | Sierena, Corcovado, Puntarenas | 13- | Concepción, Puriscal, San José |
| 0011 | San José | 14- | Jilgueral de Puriscal, San José |
| 10354 | No data | 15- | Jilgueral de Puriscal, San José |
| 0102 | San Miguel de Sarapiquí, Heredia | 16- | Alto de Limón, Puriscal, San José |
| 7185 | San Clemente, Limón | 17- | Alto de Limón, Puriscal, San José |
| 2938 | Penshurst, Limón | 18- | La Gloria de Puriscal, San José |
| 2939 | Penshurst, Limón | 19- | La Gloria de Puriscal, San José |
| 2740 | Penshurst, Limón | 20- | La Gloria de Puriscal, San José |
| 2741 | Penshurst, Limón | 21- | Talarnanca, Río Telire, Limón |
| 10306 | Pilón de Bijagua, Upala, Alajuela | 22- | Turrubares, San José |
| 2212 | La Selva, Sarapiquí, Heredia | 23- | Los Angeles de Puriscal, San José |
| 11471 | Sirena de Corcovado, Puntarenas | 24- | Los Angeles de Puriscal, San José |
| 6768 | Bri-bri, Limón | 25- | Salitrales de Puriscal, San José |
| 6233 | Turrubares, San José | | |
| 3400 | La Bomba, Limón | | |
| 3405 | Puerto Vargas, Limón | | |
| 2895 | Siquirres, Limón | | |